# Understanding the Impact of Heatwave on Urban Heat Island in Greater Sydney: Temporal Surface Energy Budget Change with Land Types

Jing Kong[1], Yongling Zhao[2], Dominik Strebel[2], Kai Gao[3], Jan Carmeliet[2], Chengwang Lei[1]

1. Centre for Wind, Waves and Water, School of Civil Engineering, The University of Sydney, Sydney, Australia

2. Department of Mechanical and Process Engineering, ETH Zürich, Zürich, Switzerland

3. The Chinese University of Hong Kong, Hong Kong, China

## Abstract

The impact of heatwaves (HWs) on urban heat island (UHI) is a contentious topic with contradictory research findings. A comprehensive understanding of the response of urban and rural areas to HWs, considering the underlying cause of surface energy budget changes, remains elusive. This study attempts to address this gap by investigating a 2020 HW event in the Greater Sydney Area using the Advanced Weather Research and Forecasting (WRF) model. Findings indicate that the HW intensifies the nighttime surface UHI by approximately 4°C. An analysis of surface energy budgets reveals that urban areas store more heat during the HW due to receiving more solar radiation and less evapotranspiration compared to rural areas. The maximum heat storage flux in urban during the HW can be around 200 W/m$^2$ higher than that during post-HW. The stored heat is released at nightime, raising the air temperature in the urban areas. Forests and savannas have relatively lower storage heat fluxes due to high transpiration and albedo, and the maximum heat storage flux is only around 50 W/m$^2$ higher than that during post-HW. In contrast, a negative synergistic effect is detected between the 2-m UHI and HW. This may be because other meteorological conditions including wind have substantial impacts on the air temperature pattern. The strong hot and dry winds coming from the west and the proximity of tall buildings to the coast diminish the sea breeze coming from the east, resulting in a higher air temperature in the western urban district. Meanwhile, the western forest area also experiences higher temperatures due to the westward winds. In addition, changes in wind direction alter the temperature distribution in the northern rural region. Based on the present study, urban climate simulation data and associated findings can be used to develop urban heat mitigation strategies for UHI during HW.

# 1 Introduction

Urban heat island (UHI) occurs and evolves as a result of urbanization in which rural areas are often replaced by buildings and roads and in turn surface characteristics, such as albedo, vegetation fraction and roughness length are modified [1-3]. These alterations affect the land surface-atmosphere interactions, causing modifications to heat flux exchange and regulating the partitioning of available energy into sensible, latent and storage heat fluxes, ultimately resulting in altered surface and air temperatures [1, 4-6]. Chen et al. [7] investigated the land cover change in Wuhan (China) from 2007 to 2020 with urban expansion increasing from 32.81% to 46.01%, and discovered that the average land surface temperature rose from 25.92 °C to 31.71 °C. Li et al. [1] found that conversion from croplands (19.77 km$^2$) or grasslands (27.46 km$^2$) to urban and built-up land could result in an increase of air temperature by 0.23 °C during summer in the plateau area.

Heatwaves (HWs) are extreme heat events over consecutive days and may adversely impact the economy, agriculture, society, and human health. Due to increased greenhouse gas emissions, HWs are anticipated to occur more frequently, be more intense, and last longer, causing higher morbidity and mortality [8-12]. HWs may also exacerbate the UHI intensity by modifying sensible and latent heat fluxes and affecting wind speed [13-16], thereby garnering increased attention in recent years.

Many studies have reported a positive synergistic effect between UHI and HW [13-15, 17-27]. For instance, Unger et al. [20] studied the UHI effect in Szeged (Hungary) and found that the UHI intensity could be 3 times or more intense during HW. Rogers et al. [24] studied three extra-tropical coastal Australian cities of Melbourne, Adelaide, and Perth from Jan 1995 to March 2014. They reported that the nighttime UHIs in Melbourne and Adelaide were exacerbated during HW by up to 1.4°C and 1.2°C respectively. Founda and Santamouris [13] showed that 5 HW episodes in 2012 in the coastal city Athens (Greece) amplified the UHI intensity by up to 3.5°C. In other coastal cities, Shanghai (China) and New York City (US), the UHI intensity were also reported to be enhanced during HWs. Ao et al. [21] and Jiang et al. [28] found that the UHI intensity in Shanghai increased by around 1°C during HW. A similar magnitude of UHI amplification was observed by Ramamurthy and Bou-Zeid [27] in New York City during the 2016 HWs.

In contrast, other studies reported no or even a negative synergistic effect between UHI and HW [11, 29-33]. For instance, Khan et al. [30] found that in Sydney (Australia), the air temperature in rural areas was higher than that in urban areas and the difference became more pronounced during HWs. Chew et al. [11] reported that the UHI intensity in Singapore was not amplified, remaining at its peak of near 2.5 °C during HW.

Key meteorological factors which contribute to the UHI and HW synergy have been explored in various studies, but there is no clear consensus due to location-specific variations. The increase in incoming shortwave radiation due to clear sky conditions and enhanced anthropogenic heat flux during HWs were reported to be important contributing factors to the positive synergy [15, 34], while others found that decreasing evapotranspiration in rural areas due to soil moisture depletion could lead to negative synergy during HWs [31, 32].

Nevertheless, there is a disproportionate lack of studies that quantifies surface energy fluxes in different land types when investigating the synergistic effect. Land types have a significant impact on the near-surface temperature [35], and understanding energy fluxes change in different land types during HWs will provide a fundamental basis for future land use planning and UHI mitigation. To fill this research gap, this study aims to examine the impact of HW on UHI by quantifying the surface energy budget change of different land types in a coastal city - the Greater Sydney Area. The specific objectives of the present study are: (1) comparing the UHI intensity among pre-HW, HW and post-HW periods; (2) quantifying the temperature change of different land types; and (3) identifying the main drivers contributing to UHI intensity change during HWs.

## 2    Methodology

### 2.1    Geographical Location and Climate

Greater Sydney is the capital of the state of New South Wales and the largest city in Australia, located between 149.972°E to 151.631°E and 32.996°S to 34.331°S, covering an area of approximately 12368 km$^2$ [36]. It contains the metropolitan Sydney and surrounding areas including the Blue Mountains, Wollondilly, and Central Coast. Greater Sydney is also the most populous city in Australia, with a population of approximately 5.2 million in 2021 which is projected to be around 8 million by 2056 as reported by the Greater Sydney Commission [37, 38]. Its climate is classified as a humid subtropical climate (Cfa) with warm summers and cool winters under the Köppen-Geiger climate classification [39]. Large variations of temperature among the coastal areas, inland western suburbs and Blue Mountains are observed in Greater Sydney due to the regulating effect of cooling sea breeze from the east and hot and dry wind from the west, which moderate the temperature of the different areas [40-42]. In summer, the monthly averaged daily maximum temperatures recorded at Observatory Hill in the central business district (CBD) can vary between 25.2°C and 26.0°C [43]. In comparison, the monthly averaged daily maximum temperature ranges from 22°C to 24°C in the Blue Mountains and 28°C to 30°C in Western Sydney [44]. Nevertheless, due to climate change Greater Sydney has experienced an increase in temperature and more hot days with maximum temperatures above 38°C over the past several decades [43, 44].

## 2.2 WRF Model Description and Validation

The WRF model, which is developed by the National Center for Atmospheric Research (NCAR), is a state-of-the-art, fully compressible, and non-hydrostatic regional climate model [45, 46]. It has been adopted to conduct a variety of operational forecasting and demonstrated the ability to successfully extract regional climate characteristics, including UHI during HW periods [11, 14, 47-52]. We use WRF V4.3 with the Advanced Research WRF dynamics core to study the impacts of HW on UHI in Greater Sydney.

The entire model consists of three one-way nested domains, as shown in Figure 1. Since a high spatial resolution is crucial for depicting land surface heterogeneity and providing accurate data for analysis [1, 53, 54], the innermost domain D03 contains 720×720 grids with a 0.25-km resolution (domain size 180×180 km$^2$), nested into coarser outer domains D02 containing 500×500 grids with a 1.25-km resolution (domain size 625×625 km$^2$) and D01 containing 300×300 grids with a 6.25-km resolution (domain size 1875×1875 km$^2$). It should be noted that the horizontal resolution of D03 (less than 1 km) is in a gray zone in the planetary boundary layer (PBL) modelling, implying that heat fluxes and momentum in the PBL are partially resolved, in contrast to conventional PBL parameterizations that are unable to resolve any turbulence [11, 55]. Other studies using a comparable fine sub-kilometre grid have shown that the simulation results accord well with the weather station data [11, 47]. Each domain has 65 vertical levels with denser levels close to the ground surface in order to achieve a finer resolution at the lowest levels. The model simulation period is from January 1, 00:00 to February 29, 23:00, 2020 (UTC) with the first day as the spin-up time to prevent inconsistency between internal dynamics and external forcing. The simulated results are exported hourly.

The input geographical data for D01 and D02 is the moderate resolution imaging spectroradiometer (MODIS) 15 s land use data. As we focus on the output results from D03, the geographical data in D03 is modified by a 250-m high-resolution spatial land use dataset obtained from the New South Wales department of planning and environment (DPE) [56]. The original dataset from DPE contains 32 land use categories, and we subsequently use Quantum Geographic Information System (QGIS) to reclassify them into 20 land use categories in MODIS to obtain a 250-m resolution dataset. The initial and boundary meteorological conditions are 3-hourly 0.25°×0.25° ECMWF Reanalysis v5 (ERA5) model level data. Due to the long-running period of the WRF model and Sydney's coastal location, the time-varying sea surface temperature (SST) is employed to improve modelling accuracy [57].

Here we use the Noah land surface model (LSM) [2] which adopts a bulk approach and treats urban areas as homogeneous surfaces to provide skin temperatures and surface heat fluxes. The WRF model coupled with Noah-LSM has been demonstrated to reveal satisfactory performances for UHI simulation [3, 58, 59]. Mellor-Yamada-Janjić (MYJ) scheme [60], which leverages the 1.5-order closure turbulence model of Mellor and Yamada [61] to depict turbulence above the surface, is employed for PBL modelling. Other physical schemes employed are Rapid Radiative Transfer Model for GCMs (Global Circulation Models) (RRTMG) scheme [62] for longwave and shortwave radiation, new Tiedke scheme [63, 64] for cumulus, Thompson et al. scheme [65] for microphysics, and Monin-Obukhov (Janjic Eta) similarity scheme [66] for surface layer parameterization.

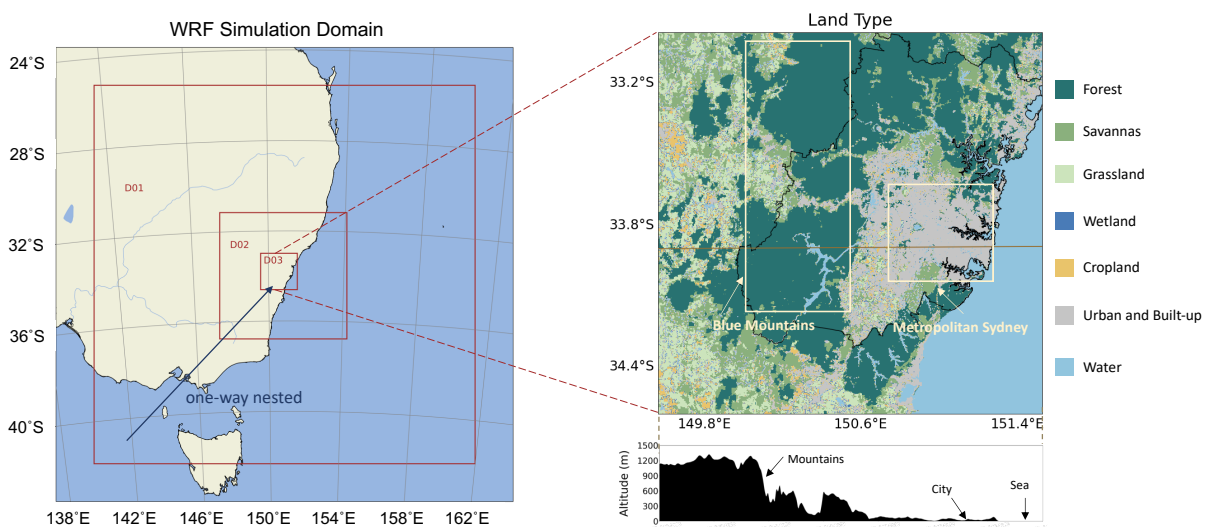

Figure 1. Configuration of the WRF input domains and land classifications in D03.

To evaluate the model performance, hourly near-surface temperature data averaged from one-minute data is used. The one-minute real-time data is obtained from a total of 8 weather stations across Greater Sydney, which are provided by the Bureau of Meteorology (BOM) (http://www.bom.gov.au). The quality evaluation indicators provided by the BOM show that the data's precision is adequate.

The following statistical indices, including the bias, root mean square error (RMSE) and Pearson's correlation coefficient (R), are employed to evaluate the discrepancy between the weather station observation and the WRF outputs from corresponding simulated grid cells:

$$Bias = (\bar{M} - \bar{O}) \tag{1}$$

$$RMSE = \sqrt{\frac{1}{N}\sum_{i=1}^{N}(M_i - O_i)^2} \tag{2}$$

$$R = \frac{\sum(M_i - \bar{M})(O_i - \bar{O})}{\sqrt{\sum(M_i - \bar{M})^2 \sum(O_i - \bar{O})^2}} \tag{3}$$

where $O$ is the weather station observation data, $M$ is the WRF model output, $N$ is the number of hourly data during the simulation period over which the calculation is made, and $i$ ranges from 1 to $N$.

Figure 2 shows the comparison between the WRF simulation and the observed data at the observatory hill weather station. It is seen in this figure that the WRF simulation captures the diurnal temperature variation well, and the overall performance of the WRF model is deemed satisfactory. More comparisons with observed data at other weather stations can be found in Appendix A.

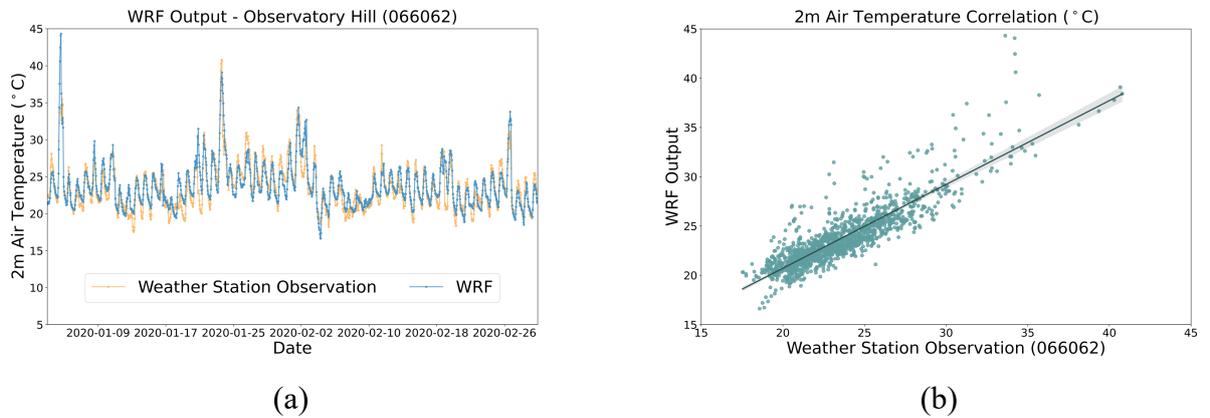

(a)                                                         (b)

Figure 2. (a) Comparison of 2-m air temperature observed at the observatory hill weather station with WRF simulation results for the period from 2 January to 28 February 2020. (b) Correlation of 2-m air temperature between WRF output results and weather station observation.

Table 1 summarises the statistics from the comparison between WRF simulation results and the observations at different weather stations. It is seen in the table that the bias is less than 1 °C and mostly positive showing WRF overpredicts the temperatures mainly during the night. The RMSE values range from 1.17 to 2.03 °C, with R values greater than 0.82. According to Ramamurthy et al. [25], WRF simulations with a 2 °C RMSE are sufficiently accurate to characterise the dynamics of excessive urban heat in cities, given the various parameterizations adopted and their complex form. The present evaluation demonstrates an accuracy comparable to or even better than other studies [47, 67-69]. Therefore, the WRF model along with the above-described setup may be adopted with confidence to resolve the spatiotemporal variation of meteorological variables and study the effect of HW on UHI.

Table 1. Bias, RMSE and R of the simulated 2-m air temperature against in-situ observations

|            | Mangrove Mountain | Sydney Airport | Terrey Hills | Observatory Hill | Holsworthy Aerodrome | Sydney Harbour | Sydney Olympic Park | Horsley Park |
|------------|-------------------|----------------|--------------|------------------|----------------------|----------------|---------------------|--------------|
| Bias (°C)  | -0.06             | 0.41           | 0.44         | 0.18             | 0.95                 | -0.15          | 0.59                | 0.70         |
| RMSE (°C)  | 2.03              | 1.66           | 1.77         | 1.59             | 2.00                 | 1.17           | 1.82                | 1.95         |
| R          | 0.91              | 0.89           | 0.90         | 0.87             | 0.92                 | 0.82           | 0.91                | 0.93         |

## 3    Results

We focus on one HW event that occurred from late January to early February 2020. Details of the HW event and the process of identifying HWs are described in Appendix B. The subsequent analysis of the results refers to three time periods: pre-HW (January 23 to 29), HW (January 30 to February 3), and post-HW (February 4 to 10). As suggested by Lo et al. [70] and Oke [71], daytime and nighttime UHI are different due to variations in energy budget change. The land surface absorbs solar radiation during the daytime and releases the stored heat during the nighttime. Therefore, all output parameters, including surface temperature, 2-m air temperature, and 10-m wind speed and direction are averaged for daytime (6 am-8 pm) and nighttime (9 pm-5 am) based on the sunrise and sunset time in Sydney summer over the three selected periods.

### 3.1   Surface Temperature

Figure 3 shows the daytime and nighttime surface temperature maps during the pre-HW, HW, and post-HW periods. Overall, the daytime and nighttime surface temperatures in both urban (mostly covered by urban and built-up) and rural (mostly covered by forests, savannas, grasslands, and croplands) areas during HW days are higher than those during non-HW days. The post-HW is a cooler period showing much lower surface temperatures compared to pre-HW and HW periods. During HW days, the highest averaged daytime urban surface temperature reaches approximately 39 °C. There is only slight difference in the surface temperature distribution patterns among these three periods. Referring to the land type map in Figure 1, the western urban part is consistently warmer than other parts over all periods, while cooler regions are present in the rural areas, especially the western part of the rural areas. The surface temperature differences between the urban and rural areas are large, indicating that the urban plays a significant role in the surface temperature change. In the urban areas, a relatively lower surface temperature is observed in the regions adjacent to the ocean, indicating distinct temperature gradients over the city. Similar results are also observed by Hirsch et al. [49] during other HW periods. The contrasting daytime and nighttime surface temperatures between the nearby ocean and the urban interior are more evident during HW days. In rural areas, the Blue Mountains area at high altitudes has the lowest surface temperature.

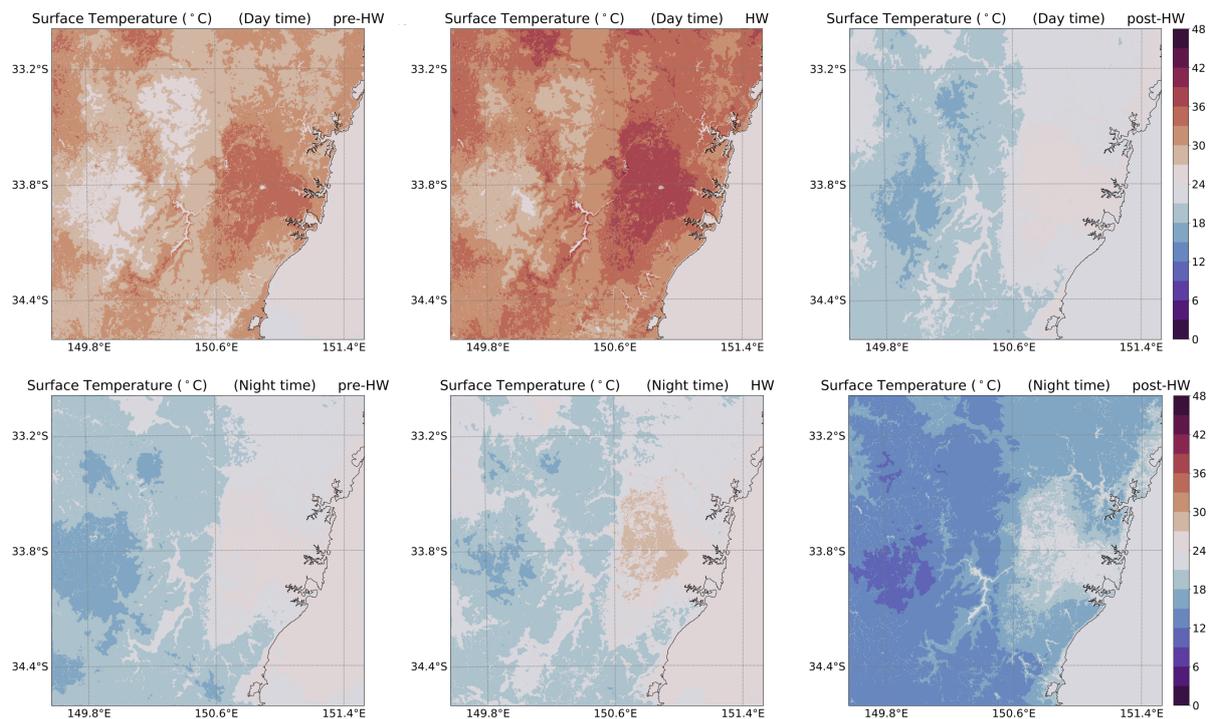

Figure 3. Averaged daytime (upper row) and nighttime (lower row) surface temperature during pre-HW (left column), HW (middle column), and post-HW (right column).

## 3.2    2-m Air Temperature and 10-m Wind

The averaged 2-m air temperature and 10-m wind fields during the three periods are shown in Figure 4. Compared with the surface temperature distribution in Figure 3, the 2-m air temperature is lower with a maximum averaged temperature around 33 °C. The western urban part is still the warmest area, but the urban heat island effect is weak. During the daytime, the differences in 2-m air temperature over the urban areas are distinctly higher during HW days than those during non-HW days. A possible explanation for this observation is related to wind effects. As depicted in the figure, the sea breeze brings cooler air to the eastern urban area, while the hot and dry wind blowing from the west potentially raises the air temperature in the western urban part. When HW arrives, the wind coming from the west becomes stronger and offsets the effect of the sea breeze in the western urban part. A heat dome is then formed over the western urban area, resulting in higher air temperatures there. During post-HW days, the intensity of the sea breeze significantly increases which brings a precipitous drop in air temperature. It is interesting to note that during the HW episode, the high air temperatures that prevail throughout the daytime in the western urban region are also present in the northern rural region. This could be because the wind direction previously originating from the southeast shifts to the northeast. Therefore, the northern rural section, which is located at a significant distance from the shore, is subject to the influence of the wind coming from the northern continent rather than the sea breeze.

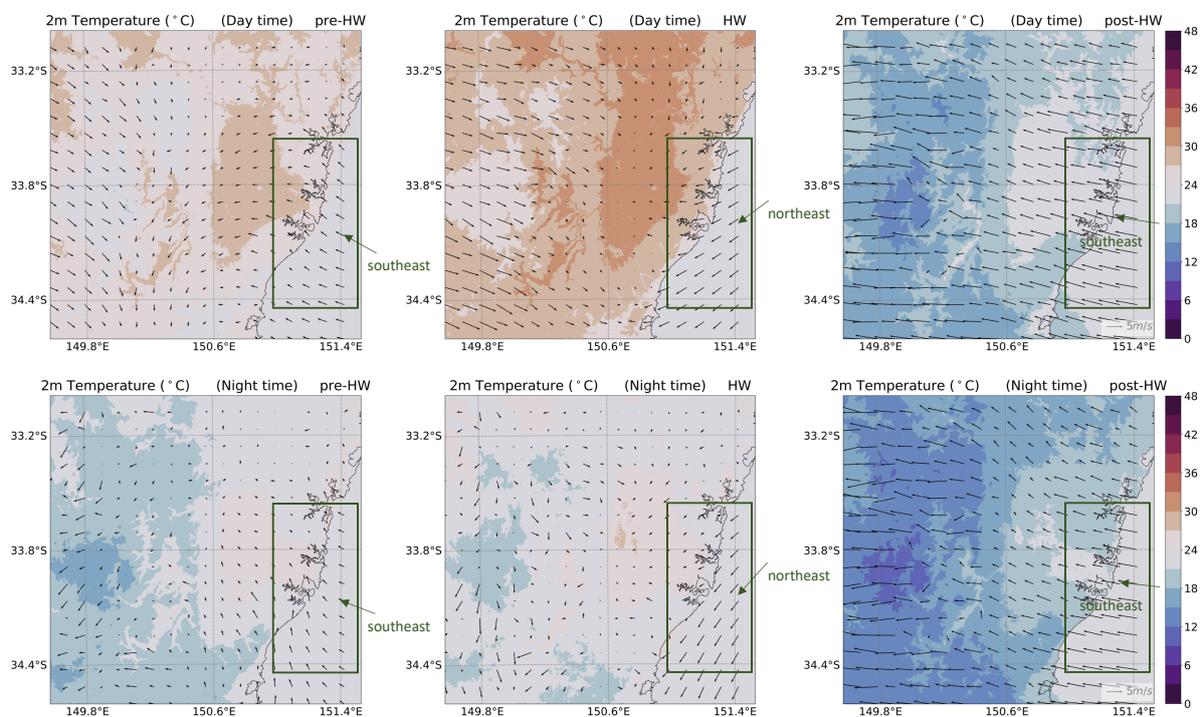

Figure 4. Averaged daytime (upper row) and nighttime (lower row) 2-m air temperature and 10-m wind field during pre-HW (left column), HW (middle column), and post-HW (right column).

## 3.3 Urban Heat Island Intensity

To avoid the altitudinal impact on air temperature, we only consider the rural areas with altitude variations within 30m relative to the main urban areas, as suggested by Martin-Vide et al. [72]. The area of different land types is shown in Table 2. Clearly, urban, forests and savannas are the three most covered land types in the Greater Sydney Area and thus will be focused on in subsequent discussions. As the thermophysical properties of forests and savannas such as albedo are different, they will be investigated individually. The grasslands, croplands and wetlands will not be considered due to their minor proportion.

Table 2. Summary of land use in the Greater Sydney Area where the altitude difference between rural and major urban areas is less than 30m.

| Land Type | Urban | Forests | Savannas | Grasslands | Croplands | Wetlands |
|---|---|---|---|---|---|---|
| Percentage (%) | 28.69 | 45.07 | 19.83 | 4.52 | 1.66 | 0.23 |

The UHI intensity is calculated by taking the average temperature differences between the urban and rural (forests and savannas) in Table 2. Figure 5(a) compares the hourly surface UHI intensities during the pre-HW, HW and post-HW periods. Overall, the results indicate a negative synergy between the surface UHI and HW during the daytime and a positive synergy during the nighttime (particularly at 0-6am). The diurnal profiles of the surface UHI intensity are comparable over all three time periods, with the highest surface UHI intensities occurring during the nighttime and the lowest or even zero surface UHI intensities occurring during the daytime. The peaks that occur during pre-HW and HW are similar, reaching approximately 4 °C, while the peaks that occur during post-HW days reach approximately 3.5 °C.

The hourly 2-m UHI intensities change is shown in Figure 5(b). In general, the 2-m UHI intensity is substantially lower than the surface UHI intensity, with the maximum and minimum values reaching around 1.5 °C and -1°C, respectively. The diurnal profile of 2-m UHI intensity is similar to the surface UHI intensity, with the UHI effect being more pronounced during the nighttime. It is interesting to note that the time when the 2-m UHI intensity achieves the minimal values is delayed compared with the surface UHI intensity, particularly during the HW period (changes from 9 am to 16 pm). Similar results have been reported for Beijing, China [15] when the 8-m and 32-m UHI are compared. However, the delay time is only around 1 hour. Negative synergy between 2-m UHI and HW are observed over the daytime when compared with that in post-HW days. When compared with pre-HW days, positive synergy is observed only between 0 and 8 am.

To investigate the impact of land type on the variation of UHI intensity, the average temperature difference between urban and two main rural land types is shown in Figure 5(c). During the daytime, the surface temperature difference between the urban areas and forests is greater than that between the urban and savannas, indicating that surface temperatures in the forests are consistently lower than those in the savannas, while opposite results are observed for the 2-m temperature. Despite the temperature spike during HW, the difference in the surface temperature and 2-m air temperature between urban and two rural land types is less, and the negative value of -0.25 °C indicates that the 2-m temperature in forests is even higher than that in urban areas. During the nighttime, the positive values of the temperature difference show that the surface and air temperatures in the savanna areas are lower than that in the forest areas over all three periods. The surface and 2-m air temperature difference between urban and two rural land types is greater than that during the daytime, with the maximum value reaching 3.67 °C. A higher surface temperature difference is observed between urban and both rural land types during HW periods while the 2-m temperature difference shows opposite results.

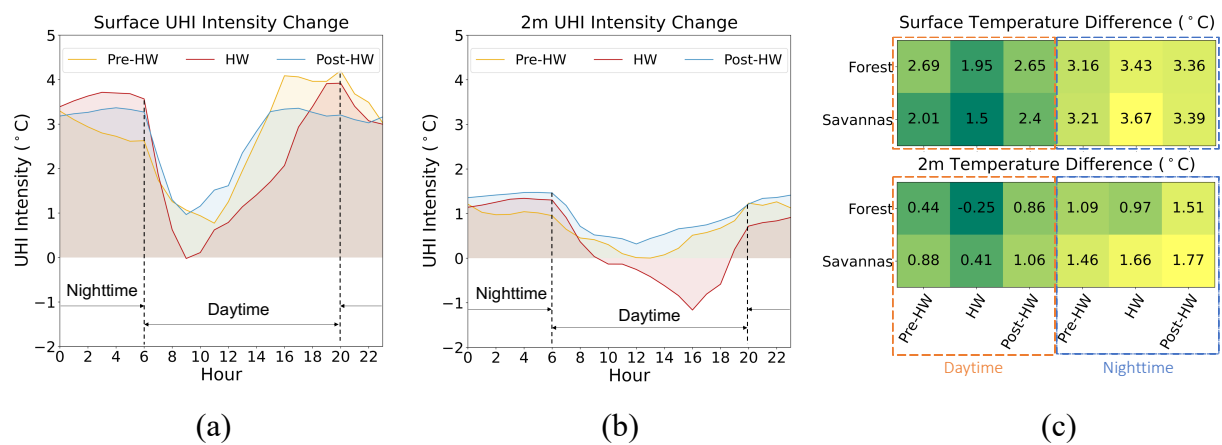

Figure 5.  (a) Hourly surface urban heat island intensity ; (b) hourly 2-m urban heat island intensity; (c) surface and 2-m temperature difference between urban and two rural land types (forests and savannas) for pre-HW, HW and post-HW periods.

## 4    Discussion

This section will focus on the surface energy budget change and other factors underlying the interactions between UHI and HW. Temporal surface energy budget change will be analysed for different land types, and the main determining factors will be revealed.

### 4.1    Surface Energy Budgets Change

The alteration of the energy balance resulting from urbanization is the root cause of UHI [71, 73], and the variation of the heat fluxes between urban and rural areas influences the UHI intensity. The surface energy budget can be expressed as [11, 15]:

$$Q^* + Q_F = Q_H + Q_E + \Delta Q_s \tag{4}$$

where $Q^* = Q^\downarrow - Q^\uparrow$, $Q^\downarrow$ is the combined downward longwave and shortwave radiation, $Q^\uparrow$ is the combined upward longwave and shortwave radiation, and $Q_F$ is the anthropogenic heat flux. $Q_F$ is not included in the following discussion since it has not been considered in the WRF simulations due to the difficulty in accurately estimating anthropogenic heat flux. $\Delta Q_S$ takes into account the heat transferred into both the ground and buildings.

Figure 6 presents the spatially and temporally averaged diurnal changes in downward radiative flux $Q^{\downarrow}$, upward radiative flux $Q^{\uparrow}$, net radiative flux $Q^*$, sensible heat flux $Q_H$, latent heat flux $Q_E$, and storage heat flux $\Delta Q_S$ during pre-HW, HW and post-HW periods, respectively. All fluxes are obtained for each land type by averaging spatially within 30 m altitude variations and temporarily over one hour.

In general, $Q^{\downarrow}$, $Q^{\uparrow}$, and $Q^*$ increase during the HW period for both urban, forests and savannas compared to non-HW periods, as shown in Figure 6 (a), (b) and (c). The increases occur throughout the daytime with a maximum at noon, which is consistent with the fact that HW days are often clear and cloudless days due to high pressure [49, 74, 75]. The comparison of cloud fractions among different periods also demonstrates that there are less clouds during HW throughout both the day and night, as shown in Appendix C. During the HW period, $Q^*$ is higher by around 50 W/m² than that in the pre-HW period (Figure 6 (c2)), and it is 300 W/m² larger than that in the post-HW period (Figure 6 (c1) and 6 (c2)). Similar results are also reported in other cities including Singapore (170 W/m² higher in HW compared to non-HW) [11] and Shanghai, China (128 W/m² higher in HW compared to non-HW) [21]. There is little difference of $Q^{\downarrow}$ among the three land types, while for $Q^{\uparrow}$, the difference among the three land types is distinct. Savannas reflect more incident radiation than urban surfaces and forests due to their higher albedo [76] as shown in Appendix D. The predominant forest type in the Greater Sydney Area is the evergreen broadleaf forest, whose albedo is low by roughly 0.1 compared to that of the savannas and urban surfaces, thus reflecting less radiation. Combing $Q^{\downarrow}$ and $Q^{\uparrow}$ together, the response of $Q^*$ to HW in various land types reveals pronounced differences. The total radiation absorbed by savannas is significantly less than that absorbed by urban areas and forests (Figure 6 (c)), which is one of the reasons why the temperature in savannas is lower.

$Q_E$ released into the atmosphere is associated with soil evaporation and plant transpiration [15]. Since urban areas are mostly covered with impervious surfaces, $Q_E$ in the urban areas remains close to 0 and is significantly lower than that in the forests and savannas during the daytime, as shown in Figure 6 (d). $Q_E$ in the forests is the highest, and its maximum value, which is around 300 W/m², is approximately twice as high as that in savannas (Figure 6 (d2) and 6 (d3)). This may explain the lowest daytime surface temperature in the forests. While during the nighttime, $Q_E$ is comparable in different land types and is nearly 0 W/m². During HW days, $Q_E$ increases in the forests indicating stronger transpiration due to increased incident radiation, whereas in savannas, $Q_E$ remains constant. The results are similar to those observed in Singapore [11]. $Q_E$ is slightly higher than 0 W/m$^2$ in commercial/industrial and high-density residential areas while it is around 350 W/m$^2$ in broadleaf forests [11].

$Q_H$ is the primary contributor to atmospheric heating. It peaks at noon for urban and rural areas during HW and non-HW days, as shown in Figure 6 (e). In the urban areas, $Q_H$ can be as high as approximately 470 W/m$^2$, while the maximum value of $Q_H$ is around 370 W/m$^2$ for the forests. The urban $Q_H$ remains positive throughout the day due to heat storage released at night, while $Q_H$ in the forests and savannas is slightly below 0 W/m$^2$ at nighttime, as shown in Figure 6 (e1) and 6 (e2). Similar observations are also observed in Shanghai, China [21] and Singapore [11]. These observations are different from that reported for Beijing, China [15] where $Q_H$ in the urban is lower than 0 W/m$^2$ at nighttime. No remarkable changes of $Q_H$ during HW are observed in either urban or rural areas when compared with that during pre-HW, while $Q_H$ shows a considerable downward trend during post-HW. The maximum difference can reach around 150 W/m$^2$ in the afternoon.

Figure 6 (f) shows that $\Delta Q_s$ increases during the daytime when HW arrives, and it varies greatly in urban areas. Studies conducted by Sun et al. [77] for Beijing, Lodz, London and Swindon also revealed increased $\Delta Q_s$ during HW, while Chew et al. [11] reported constant $\Delta Q_s$ during HW in Singapore. During the nighttime, negative $\Delta Q_s$ is observed in both urban and rural areas, indicating the stored heat is released at nighttime. It is also noted that $\Delta Q_s$ in the urban is much higher than that in the forests and savannas during the daytime. This is consistent with the findings reported by Grimmond and Oke [78] that $\Delta Q_s$ is a significant component of the energy balance and is higher in urban areas. Therefore, more heat is released in the urban areas during the night, resulting in the relatively stronger nocturnal UHI effect.

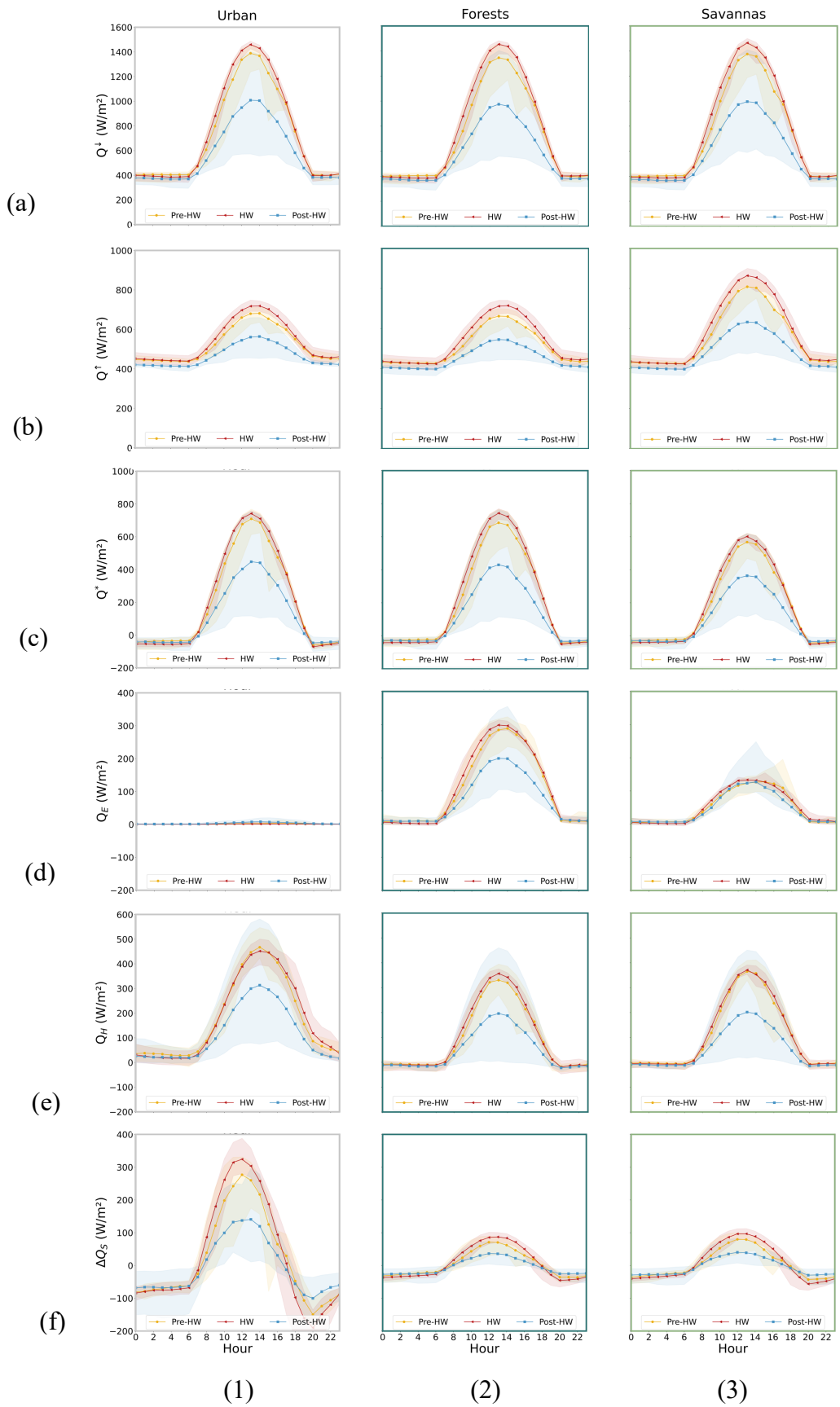

Figure 6. Hourly spatially averaged (a) downward radiative heat flux ($Q^\downarrow$), (b) upward radiative heat flux ($Q^\uparrow$), (c) net radiative heat flux ($Q^*$), (d) latent heat flux ($Q_E$), (e) sensible heat flux ($Q_H$), and (f) storage heat flux ($\Delta Q_s$) during pre-heatwave, heatwave, and post-heatwave for urban forests and savannas categories (The shaded area denote the standard deviation).

Figure 7 shows the accumulated $Q^\downarrow$, $Q^\uparrow$, $Q^*$, $Q_H$, $Q_E$, and $\Delta Q_s$ over the simulated period covering pre-HW, HW and post-HW days for different land types. The accumulated $Q^*$ (Figure 7 (c)) remains the highest in the forests and urban and is lower for the savannas. For all three land types, there are significant increases of $Q^*$ during pre-HW and HW periods, but $Q^*$ remains almost constant during post-HW, indicating much of solar radiation is absorbed during pre-HW and HW period. A similar time evolution is observed in accumulated $Q_E$ and $Q_H$ as shown in Figure 7 (d) and Figure 7 (e), respectively. The accumulated $Q_E$ is the highest in the forests, lower in the savanna and remains almost zero in urban. The accumulated $Q_H$ is the highest in urban and almost the same in savannas and forests. Interestingly, the accumulated $\Delta Q_s$ rises mostly during HW, at which point the accumulated $\Delta Q_s$ achieves its maximum value, as shown in Figure 7 (f). During post-HW, it begins to decline. The accumulated $\Delta Q_s$ in urban areas are much higher than that in rural areas, and this result explains the higher surface and air temperature in urban areas, especially during the nighttime when the stored heat is released. Similarly, the difference of accumulated $\Delta Q_s$ between urban and rural areas are greater during HW, which may explain why the nighttime surface UHI intensity is enhanced during HW.

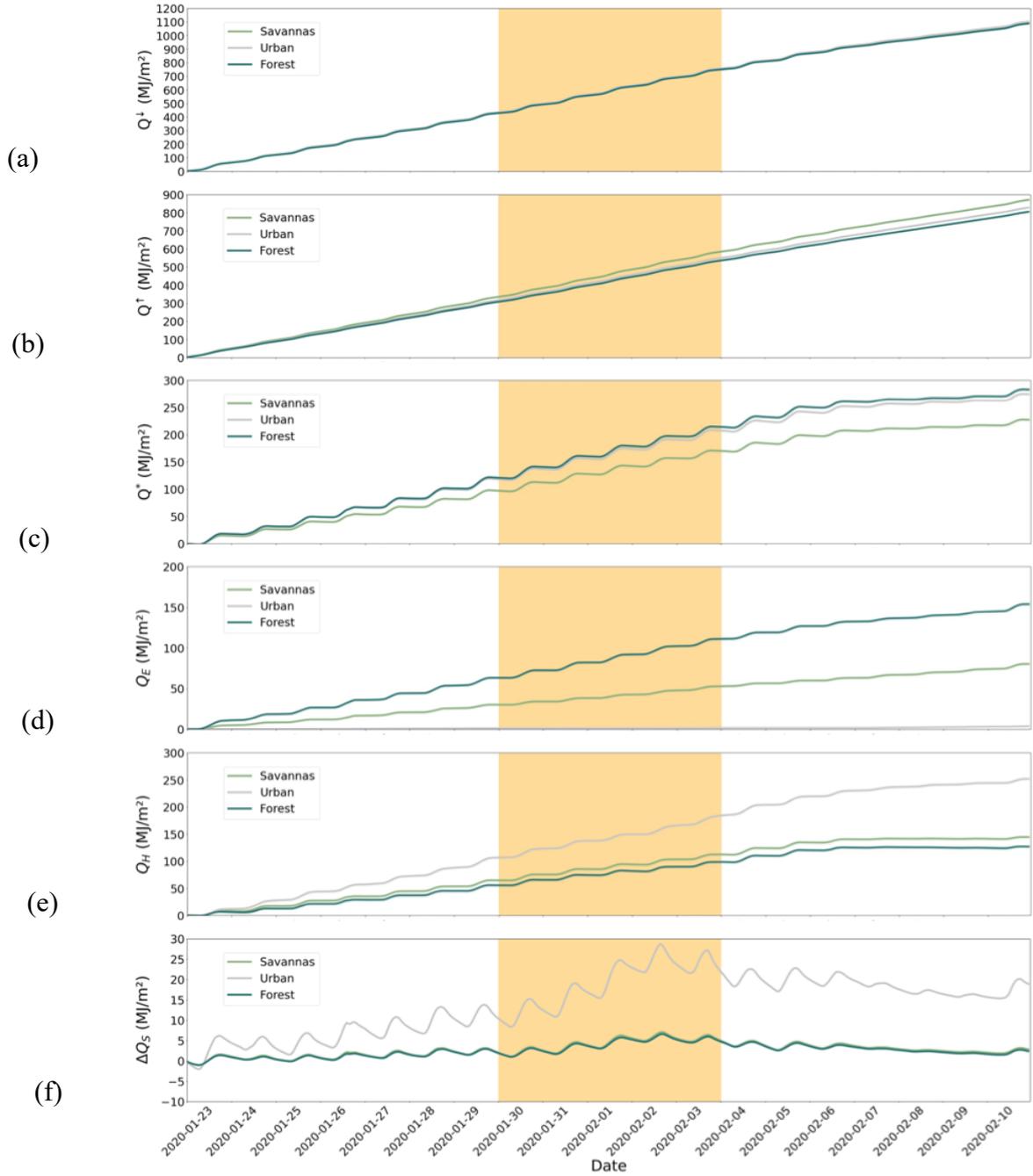

Figure 7. Accumulated (a) downward radiative heat flux ($Q^{\downarrow}$), (b) upward radiative heat flux ($Q^{\uparrow}$), (c) net radiative heat flux ($Q^{*}$), (d) latent heat flux ($Q_E$), (e) sensible heat flux ($Q_H$), and (f) storage heat flux ($\Delta Q_S$) during pre-heatwave, heatwave, and post-heatwave for urban and rural categories (forests and savannas). The yellow shaded period represents the HW days.

## 4.2   Contributing Parameters of UHI

To better understand the main drivers contributing to the interactions between UHI and HW, we study the correlations of surface temperature and 2-m temperature with other meteorological variables such as PBL height (calculated by MYJ scheme using turbulent kinetic energy as its primary diagnostic variable), 10-m wind speed, 10-m wind direction, and storage heat flux for urban, forests and savannas during pre-HW, HW and post-HW days. In a first approximation, we assume a linear relation between the variables, although nonlinear relations may exist. The correlations show quite some scatter due to high hourly variations. The correlation coefficient $r$ is shown in the lower right corner of each figure, and a higher $r$ indicates a stronger correlation between the meteorological variable and observed temperature.

It can be observed that the storage heat flux (Figure 8 (a)) and PBL height (Figure 8 (b)) have the strongest correlation with surface and air temperature with maximum $r$ reaching 0.9, followed by the 10-m wind speed (Figure 8 (c)) and direction (Figure 8 (d)). The surface temperatures in both urban and rural areas have a stronger correlation with the storage heat flux than the 2-m temperature. This finding is consistent with the higher change of the surface UHI intensity during HW compared to the 2-m air temperature since a positive synergy effect is observed at night due to more storage heat flux being released in urban areas. The PBL height has a comparable correlation with the surface and 2-m temperature. This is because the PBL height determines the effective heat capacity of the atmosphere and thus affects both the surface and air temperature pattern. Changes of the 10-m wind speed and direction regulate the aerodynamic coefficients and thus have a stronger impact on the 2-m air temperature compared to the surface temperature. During HW days, the 10-m wind speed and direction have stronger correlations with the 2-m air temperature in forest and savanna areas than that in urban areas, indicating that the forest and savanna areas are affected more by the hot and dry wind coming from the west. This partly explains the negative synergy between the 2-m UHI and HW.

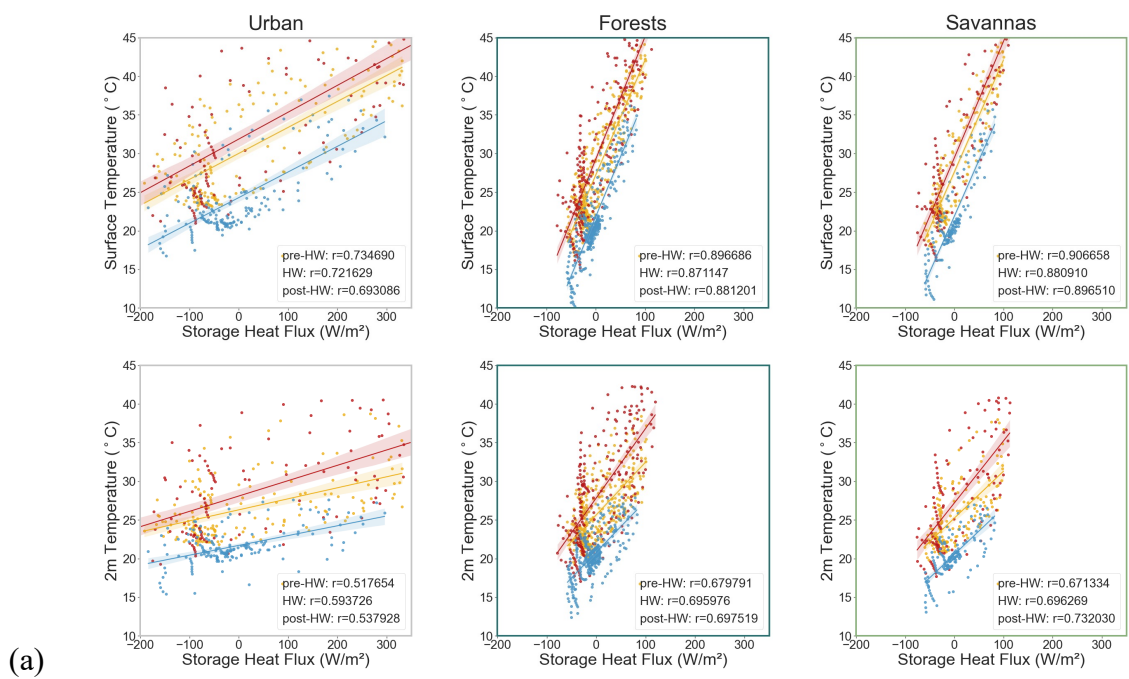

(a)

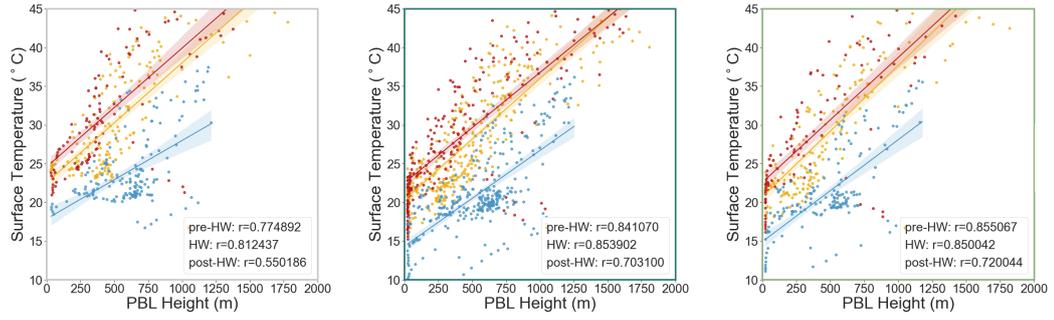

(b)

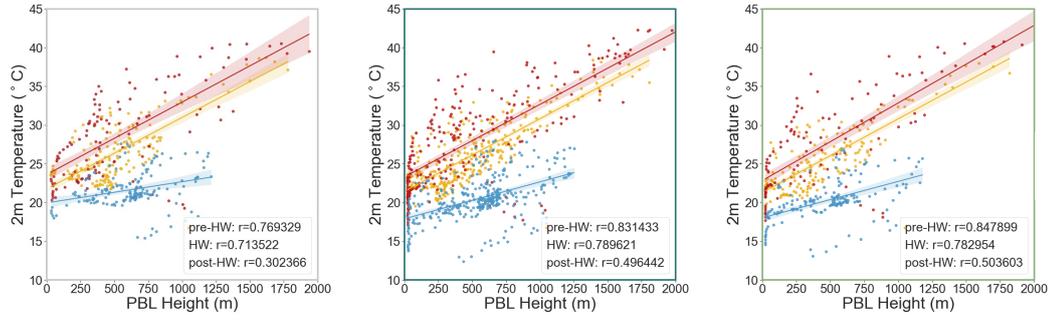

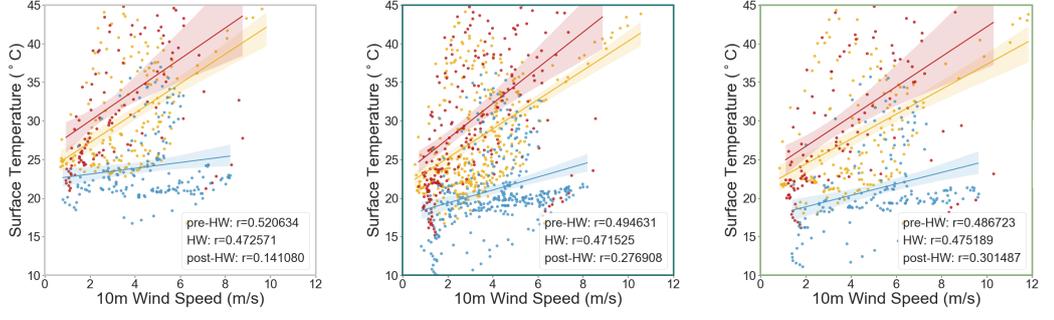

(c)

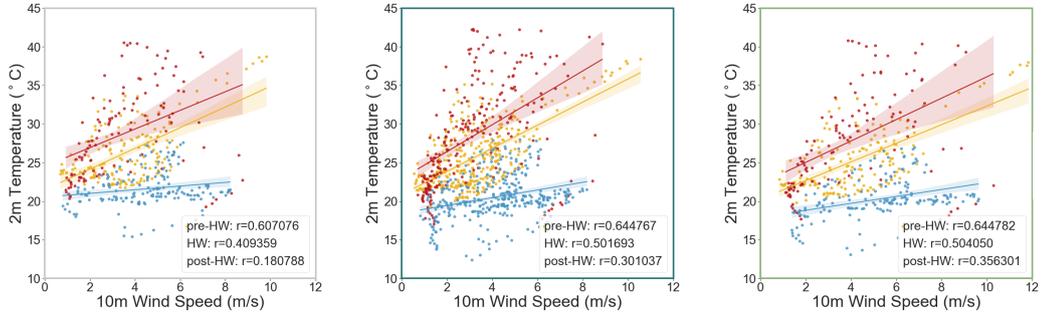

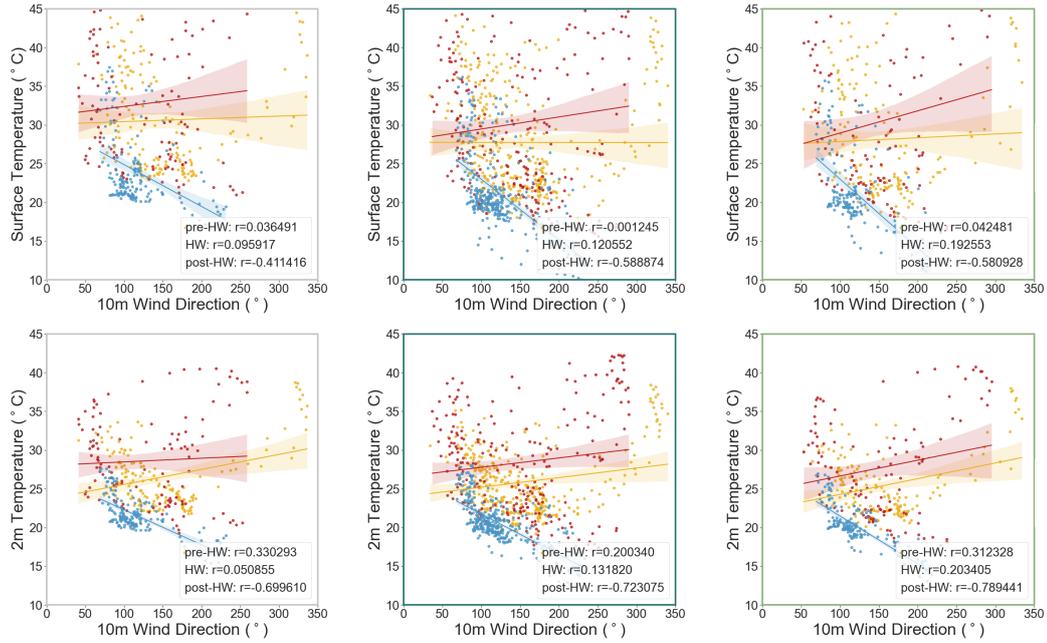

(d)

Figure 8. Correlations of the surface temperature and 2-m temperature with (a) storage heat flux, (b) PBL height, (c) 10-m wind speed, (d) 10-m wind direction in urban, forests and savannas during pre-HW (yellow), HW (red) and post-HW (blue) days.

## 5    Limitations and Future Work

This study is limited to a specific coastal city and focuses on a single HW event. Conclusions may be different for other cities and HW events. Therefore, several aspects remain to be investigated in future research. More HW events need to be considered to identify the general trend, and more cities with different urban morphologies and background climates need to be simulated to reveal the determining factors.

This study does not include the urban canopy model and thus does not consider urban inhomogeneity and variations of anthropogenic heat release. Since modelling urban climate involves strongly non-linear interactions of various processes and contributing parameters, increasing the complexity of the model does not necessarily lead to improvements in model accuracy [11, 79]. Meanwhile, more computing resources will be required with more complex models. In addition, accurately estimating anthropogenic heat flux in urban environments remains a challenge [15]. The Noah LSM is one of the most widely used land surface schemes for research and climate modelling, and its accuracy in simulating urban climates has been demonstrated in various studies [3, 80]. Since our study mainly focuses on the difference between urban and rural types and the present simulation results show a good agreement with the weather station observation data, the present simulation results are deemed credible.

# 6 Conclusions

In this study, we explore the effect of heatwave (HW) on urban heat island (UHI) based on WRF simulations of the Greater Sydney Area for an HW event in 2020. The variations of UHI and contributing parameters are analysed and compared among pre-HW (January 23 to 29), HW (January 30 to February 3) and post-HW (February 4 to 10) periods. The major findings of this study are:

- HW amplifies the surface UHI intensity at nighttime, and the peak surface UHI intensity reaches nearly 4°C. A negative synergistic effect is observed between the 2-m UHI and HW, and the minimum 2-m UHI intensity is less than 0 °C.

- The storage heat flux increases during pre-HW and HW, especially in urban areas. This is because more radiation is absorbed during HW due to cloudless conditions. Urban areas have the highest storage heat flux in the daytime since most of the urban areas are impervious surfaces, and the latent heat flux in urban areas is close to 0 W/m$^2$. Forests and savannas have relatively lower storage heat flux due to high transpiration and albedo, which result in high latent heat flux and low solar absorption, respectively.

- The storage heat flux and PBL depth have strong positive correlations with the surface and 2-m air temperatures. The greater amount of heat that is stored in urban areas during daytime and released during the night is the primary cause of the positive synergistic effect between surface UHI and HW at nighttime.

- Changes of the 10-m wind speed and direction regulate the aerodynamic coefficients and affects the temperature variation. During HW days, the strong hot and dry wind coming from the west diminishes the sea breeze from the east. Urbanization also reduces the sea breeze speed and increases aerodynamic resistance. Meanwhile, the wind direction originating from the southeast during pre-HW shifts to the northeast during HW and thus raises the air temperature in the rural northern section which is located at a significant distance from the shore. Consequently, negative synergistic effect is detected between the 2-m UHI and HW.

These findings imply that the urbanization-induced surface energy budget change and near-surface circulation play crucial roles in the intensification of UHI intensity in the event of HW. The improved understanding of the UHI and HW interactions in the Greater Sydney Area enables us to better forecast how future HWs may affect human life, allowing for the identification of appropriate mitigation strategies. In turn, this will enable policymakers to implement ways to mitigate UHI, especially during HW events.

**Acknowledgement**

The first and last authors are grateful for the financial support of the School of Civil Engineering, the University of Sydney. The first and last authors are also grateful for the numerical resources provided by the National Computational Infrastructure (NCI), Australia.

**Appendix A: Validation**

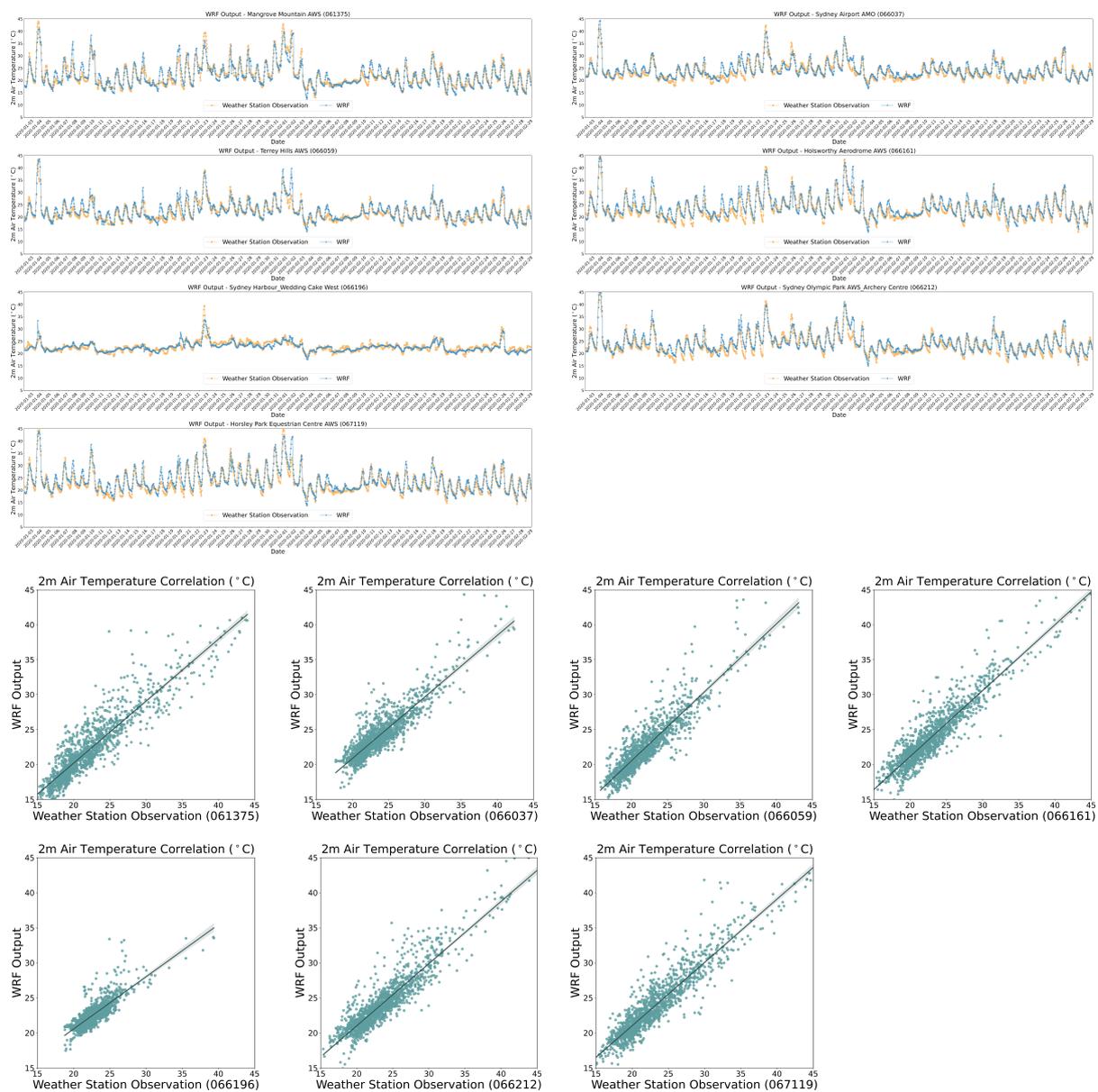

Figure A.1. Comparison of 2m air temperature between WRF output and weather station observation in different locations across the Greater Sydney Area.

**Appendix B: Heatwaves in Greater Sydney Area**

The Bureau of Meteorology defines an HW as three or more consecutive days when daytime and nighttime temperatures are extremely high compared to the local long-term climate. Here we employ the 90th-percentitle threshold based on the daily maximum temperature (daytime) and daily minimum temperature (nighttime) in summer (December - February) in recent 30 years (1991 - 2020). Two weather stations in urban areas, including Sydney Observatory Hill (066062) and Sydney Airport Amo (066037), are used to provide the data. The results are summarized in Table B.1. We choose days from 30/01/2020 to 03/02/2020 as one HW period.

Table B.1. Heat extremes in all weather stations.

| Date | Daytime (90th – 30.4°C) | | Nighttime (90th – 22.1°C) | |
|------|--------|--------|--------|--------|
| | 066062 | 066037 | 066062 | 066037 |
| 30/01/2020 | | √ | | |
| 31/01/2020 | √ | √ | √ | √ |
| 01/02/2020 | √ | √ | √ | √ |
| 02/02/2020 | √ | | √ | √ |
| 03/02/2020 | √ | | √ | |

**Appendix C: Cloud Fraction**

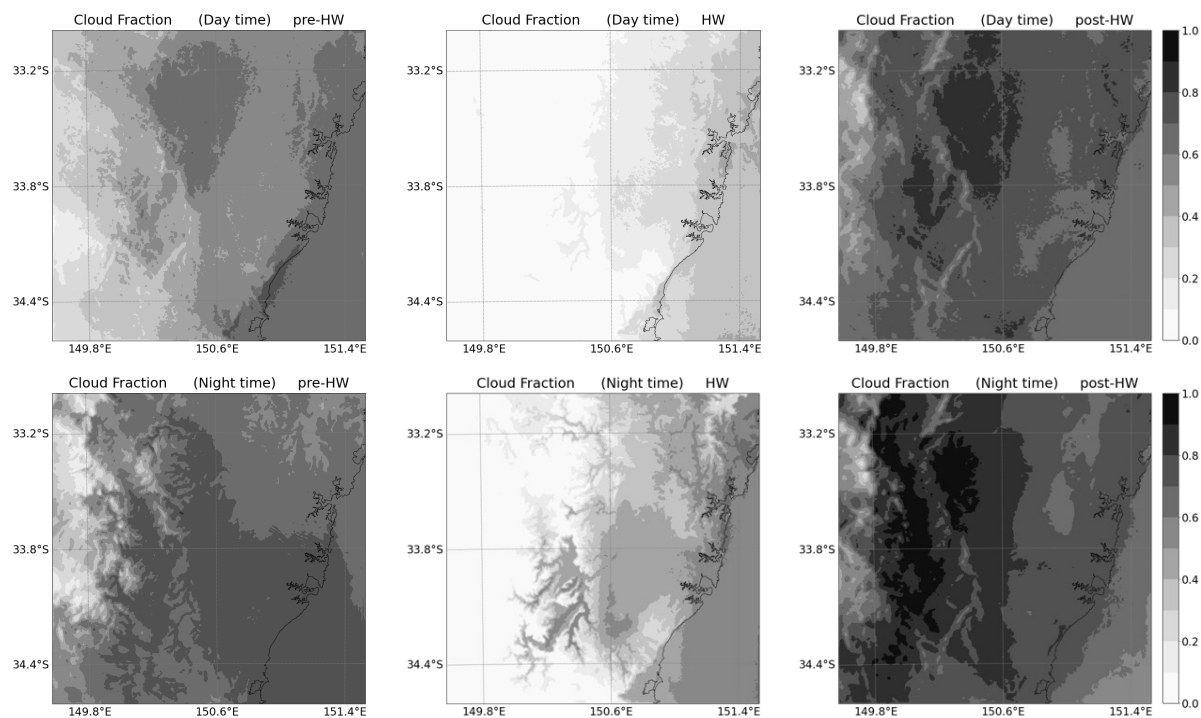

Figure C.1. Averaged daytime and nighttime cloud fraction during pre-HW, HW, and post-HW.

**Appendix D: Parameters used for simulation in WRF**

Table D.1. Parameters used for simulation in different land types in WRF.

| Land Type | Emissivity (-) | Green vegetation fraction (%) | Albedo (-) | Anthropogenic heat (W/m$^2$) |
|-----------|----------------|-------------------------------|------------|------------------------------|
| Urban | 0.88 | 0.1 | 0.15 | 20 |
| Forests | 0.94 | 0.875 | 0.14 | 0 |
| Savannas | 0.92 | 0.5 | 0.2 | 0 |

**Appendix E: Planetary Boundary Layer Depth**

Figure E.1 shows the averaged PBL depth during HW and non-HW periods. In general, the distribution pattern of the PBL depth is comparable to that of the temperature field, with the PBL being deeper in the urban region, particularly in the western urban section. During the daytime, the PBL depth in the urban region in HW days is deeper than that in non-HW days, implying higher temperatures in HW days. During the nighttime, there is no significant change of the PBL depth between the pre-HW and HW days. However, during the days following a heat wave, the PBL depth increases in both urban and rural areas.

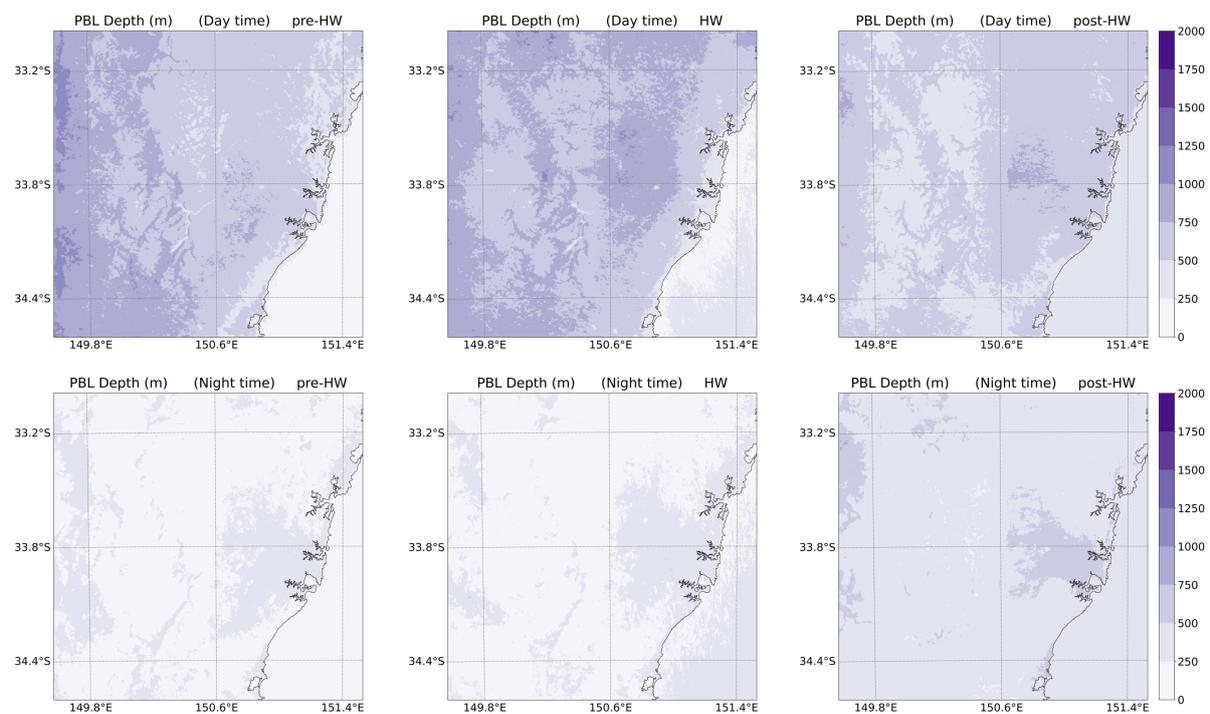

Figure E.1. Averaged daytime and nighttime PBL depth during pre-HW, HW, and post-HW.